\documentclass[preprint,number,12pt]{elsarticle}

\usepackage{amssymb}
\usepackage{amsthm}
\usepackage{amsmath}

\usepackage[left]{lineno} 
\usepackage[dvipsnames]{xcolor}
\usepackage{enumitem}

\usepackage{caption}
\usepackage{subcaption}

\usepackage[acronym]{glossaries}
\newacronym{var}{VaR}{Value-at-Risk}
\newacronym{es}{ES}{Expected Shortfall}
\newacronym{mpmr}{MPMR}{most probable maximum size of risk events}
\newacronym{emr}{EMR}{expected maximum size of risk events}
\newacronym{bm}{BM}{block maxima}
\newacronym{soc}{SOC}{self-organized criticality}
\newacronym{ti}{TI}{tail index}
\newacronym{tdc}{TDC}{tail dependence coefficient}
\newacronym{se}{SE}{scaling exponent}

\usepackage{hyperref} \hypersetup{colorlinks=true, linkcolor=blue, filecolor=magenta, urlcolor=cyan, pdftitle={Overleaf Example}, pdfpagemode=FullScreen,}

\usepackage[noabbrev]{cleveref}

\newcommand*\diff{\mathop{}\!\mathrm{d}}

\journal{The Journal of Finance and Data Science}
\let\today\relax
\makeatletter
\def\ps@pprintTitle{%
    \let\@oddhead\@empty
    \let\@evenhead\@empty
    \def\@oddfoot{\footnotesize\itshape
         {Submitted preprint} \hfill\today}%
    \let\@evenfoot\@oddfoot
    }
\makeatother

\begin{document}

\begin{frontmatter}



\title{Measuring Tail Risks}


\author[rmi,math]{Kan Chen\corref{cor1}}
 \author[rmi]{Tuoyuan Cheng}

 \cortext[cor1]{Corresponding author: kan.chen@nus.edu.sg}

 \address[rmi]{Risk Management Institute, National University of Singapore, 04-03 Heng Mui Keng Terrace, I3 Building, Singapore, 119613, Singapore}
 \address[math]{Department of Mathematics, National University of Singapore, Level 4, Block S17, 10 Lower Kent Ridge Road, Singapore, 119076, Singapore}


\begin{abstract}
\gls{var} and \gls{es} are common high quantile-based risk measures adopted in financial regulations and risk management. In this paper, we propose a tail risk measure based on the \gls{mpmr} that can occur over a length of time. \gls{mpmr} underscores the dependence of the tail risk on the risk management time frame. Unlike \gls{var} and \gls{es}, \gls{mpmr} does not require specifying a confidence level.  We derive the risk measure analytically for several well-known distributions. In particular, for the case where the size of the risk event follows a power law or Pareto distribution, we show that \gls{mpmr} also scales with the number of observations $n$ (or equivalently the length of the time interval) by a power law, $\text{MPMR}(n) \propto n^{\eta}$, where $\eta$ is the \gls{se}. The scale invariance allows for reasonable estimations of long-term risks based on the extrapolation of more reliable estimations of short-term risks. The scaling relationship also gives rise to a robust and low-bias estimator of the \gls{ti} $\xi$ of the size distribution, $\xi = 1/\eta$.  We demonstrate the use of this risk measure for describing the tail risks in financial markets as well as the risks associated with natural hazards (earthquakes, tsunamis, and excessive rainfall). 
%
\end{abstract}




\end{frontmatter}


\glsresetall
\section{Introduction}
\label{sec_1}
\noindent
Measuring tail risk is essential in quantitative risk management. There are a few established measures in financial risk management \cite{he2022risk, adam_spectral_2008}. Variance or standard deviation of returns is often used as a risk measure in the context of portfolio optimization and risk attribution. It is, however, not a good measure for tail risks as the return distributions of financial assets often exhibit fat tails. \gls{es} and \gls{var} focus on tail distributions. They are used as key metrics for determining capital requirements in financial regulations and are now widely accepted in portfolio optimization. \gls{var} and \gls{es} are specified with a given confidence level in a specific time frame. For example, the market risk is specified using the loss in $10$ days with a confidence level of $99\%$ and the credit and operational losses are specified using the loss within one year with a confidence level of $99.9\%$ \cite{hull2015}. \gls{var} is more intuitive and more popular. However, it doesn’t satisfy the key sub-additivity condition and is thus not a coherent risk measure \cite{he2022risk}. This is the main reason regulators are moving towards adopting \gls{es}, which is a coherent measure. In our opinion, the main drawback of \gls{var} and \gls{es} is that the confidence level specification is rather subjective. The tail probabilities of losses over different time frames are related; so the confidence level shouldn’t be specified independently. Instead, it should be tied to the time frame considered. 
\\\\
The tail part of the size distribution can often be characterized by a power law, which expresses the scaling relationship between the probability of a risk event and its size \cite{davison2015statistics, gomes_extreme_2015, nolde_extreme_2021}. There is a vast literature on this subject. In stock markets, the cubic law describing the tail part of the stock return distribution has been established \cite{gabaix_theory_2003, gabaix_institutional_2006, stanley2000introduction, gabaix2016power}. Kelly and Jiang \cite{kelly2014tail} have further proposed a dynamic tail risk measure, making use of the cross-section of individual stock returns to estimate the common tail index for the conditional tail risk distributions of individual assets. They provide evidence that tail risk has large predictive power for aggregate stock market returns and is important for asset pricing. There is a wide range of instances of frequency-size scaling. The distribution of operational losses can be characterized by a power law \cite{deFontnouvelle2006}. The Gutenberg-Richter law in seismology is perhaps the best-known frequency-size scaling relationship. In hydrology, the scale invariance has also been used to characterize the rainfall intensity-duration-frequency relationship \cite{burlando1996scaling, ghanmi2016estimation}. The frequency-size scaling has been viewed as a sign of emerging behavior in large interactive complex systems \cite{bak1996complexity, bak1991self, sethna2022power}. One possible mechanism, \gls{soc}, has been proposed to explain the emergence of such scaling relations \cite{bak1987self}. The concept of \gls{soc} has been applied, for example, to the study of earthquake dynamics \cite{carlson1991, chen1991, rundle2000, sornette1989, chen1997}. \gls{soc} helps to explain the mechanism for the emergence of the Gutenberg-Richter law.
\\\\
In this paper, we propose a risk measure based on the \gls{mpmr} that can occur over a time frame. This measure is generic; it is applicable for describing tail risk in financial systems as well as the risks associated with natural hazards. In the context of financial risks, the size is the financial loss. When defining \gls{mpmr} for the financial risk we don’t consider the total loss over the period as in the case of \gls{var} and \gls{es}, but only the maximum of the daily losses (regarding the market risk) or the losses from individual risk events (regarding the operational risk) in the period. We will demonstrate that the dependence of the potential maximum size of risk on the time frame considered reflects the tail size distribution of individual risk events; as the sample size increases, we will likely encounter bigger and rarer events. From the risk management perspective, the maximum loss from a shock (the duration of which is normally much shorter than the risk management time frame) might be more relevant than the total loss of the entire period. It is easy to check that \gls{mpmr} satisfies the conditions of coherent risk measures. We will explore the use of \gls{mpmr} and related \gls{emr} to characterize financial losses as well as the sizes of risk events due to natural hazards.

\section{\gls{mpmr}, \gls{emr}, and their sample size dependence}
\label{sec_2}
\noindent
In this section, we introduce the \gls{mpmr} and the \gls{emr} risk metrics and analyze their sample-size dependence for several underlying distributions. 
\\\\
Let $X_1, X_2,...,X_n$ be i.i.d. random variables with density function $f_{X}(\cdot)$ and distribution function $F_X(\cdot)$ within a sampled block of size $n$. Let $S$ be the maximum in this block (block maximum) with the density function $f_S(\cdot)$, we have \cite{bouchaud_theory_2003},
\begin{equation}\label{eq:density}
\begin{aligned}
    S&=\max_{i=1,2,...,n}{X_i}
    \\
    f_S(s)&= \frac{\partial (F_X(s))^n}{\partial s}= n ~ f_X(s) ~ \left (F_X(s)\right )^{n-1}
\end{aligned}
\end{equation}
To determine \gls{mpmr} as the most probable value of $S$, denoted as $s^*$, we set $\frac{\partial f_S(s)}{\partial s}\big|_{s=s^*}=0$, which gives,
\begin{equation}\label{eq:mpmr}
\begin{aligned}
    \left[
    \frac{\partial f_X(s)}{\partial s} \left( F_X(s) \right)^{n-1} 
    +
    f_X(s)(n-1) \left( F_X(s) \right)^{n-2} f_X(s)
    \right]\bigg|_{s=s^*}
    &=0
    \\
    \left[
    \frac{\partial f_X(s)}{\partial s} \left( F_X(s) \right)
    +
    (n-1)f_X^2(s)
    \right]\bigg|_{s=s^*}
    &=0
\end{aligned}
\end{equation}
Focusing the leading order of the large sample size limit $n\rightarrow\infty, s\rightarrow\infty, F_X(s)\rightarrow1$, we have  approximately,
\begin{equation}\label{eq:root}
    \left[
    \frac{\partial f_X(s)}{\partial s}
    +n {f_X^2(s)}
    \right]\bigg|_{s=s^*}=0
\end{equation}
\noindent
\cref{eq:mpmr} and \cref{eq:root} can be solved numerically via root-finding algorithms when $f_X(\cdot)$ is specified either parametrically or non-parametrically \cite{oliveira_enhancement_2021}.
\\\\
With others held constant, $s^*$ is a function of $n$. We now explore the relationship between $s^*$ and $n$, given several well-known underlying distributions for $f_X(\cdot)$. For the power-law and exponential distributions, we have also obtained the analytical forms of the \gls{emr}, which is the expected value of the maximum.

\begin{enumerate}[label=(\roman*)]
\item
For the normal distribution, without loss of generality, let $X\sim N(0, \sigma^2)$, $~f_X(x)=\frac{1}{\sigma \sqrt{2\pi}}\exp{\left(\frac{-x^2}{2\sigma^2}\right)}$. From \cref{eq:root} we have,
\begin{equation*}
\begin{aligned}
    s^*&\approx 
    \sigma\sqrt{W_{0}\left(\frac{n^2}{2\pi}\right)}
    \\
    &\approx
    \sigma\sqrt{
    \log(\frac{n^2}{2\pi})
    - \log(\log(\frac{n^2}{2\pi}))
    }  
    \\
    &\approx
    \sigma\sqrt{2
    \log(n)}\left(1-\frac{\log(\log(n))}{4\log(n)}\right)     
\end{aligned}
\end{equation*}
\noindent
where $W_{0}(\cdot)$ is the principal branch of the product logarithm or Lambert W function, which gives the root of $x \exp{(x)}-y=0$ \cite{johansson2020computing}, and we keep the terms corresponding to the first two leading orders in the large $n$ limit. 


\item
For the exponential distribution, let $X\sim Exp(\xi)$, $f_X(x) = \xi  \exp(-\xi x)$, and $F_X(x) = 1 - \exp(-\xi x)$. From \cref{eq:mpmr} we have,
\begin{equation*}
\begin{aligned}
    -\xi^2 \exp(-\xi s^*)&(1-\exp(-\xi s^*))+(n-1)\xi^2\exp(-2\xi s^*)=0
    \\
    s^*&= 
    \frac{\log n}{\xi}
    \\
    \mathbb{E}[S]&=\int_0^{+\infty} s f_S(s) \diff{s} =\frac{\sum_{i=1}^{n}\frac{1}{i}}{\xi}
    \approx 
    \frac{\gamma+\log{n}}{\xi} 
\end{aligned}
\end{equation*}
\noindent
where $\gamma=\lim_{n\rightarrow\infty}(-\log n + \sum_{i=1}^n{\frac{1}{i}})\approx0.577216$ is the Euler's constant. 

\item
For the power-law or Pareto type I distribution, let $Y\sim Exp(\xi)$, we have $X=A \exp(Y) \sim Pareto(A,\xi)$,  $f_X(x)=\frac{\xi A^\xi}{x^{(\xi+1)}}$, and  $F_X(x)=1-\frac{ A^\xi}{x^{\xi}}$. From \cref{eq:mpmr} we have,
\begin{equation*}
-\frac{\xi(\xi+1)A^{\xi}}{s^{\xi+2}}(1-\frac{ A^\xi}{s^{\xi}})+(n-1)\frac{\xi^2A^{2\xi}}{s^{2\xi+2}}=0.
\end{equation*}
The solution ($s=s^*$) of the above equation is given as
\begin{equation}\label{eq:MPMR:Pareto}   
    s^* = 
    A {\left(\frac{1+ n \xi}{1+\xi}\right)}^{\frac{1}{\xi}}
    \approx A(\frac{\xi}{1+\xi})^{1/\xi} n^{1/\xi}(1+\frac{1}{n\xi^2})
\end{equation}
\noindent
When $\xi>1$, using Euler's reflection formula and the reciprocal gamma function \cite{schneider2018nist}, we have
\begin{equation}\label{eq:EMR:Pareto}
\begin{aligned}
    \mathbb{E}[S]
    &=\int_A^{+\infty} s f_S(s) \diff{s} 
    =\frac{A ~ \mathcal{B}(n,1-\frac{1}{\xi}) }
    {-\mathcal{B}(-n,1)}
    =An\mathcal{B}(n,1-\frac{1}{\xi})
    \\
    &\approx 
    A~ \Gamma(1-\frac{1}{\xi}) ~ n^\frac{1}{\xi} +
    \frac{A~ \Gamma(2-\frac{1}{\xi})}{2\xi} ~ n^{\frac{1}{\xi}-1} 
\end{aligned}
\end{equation}
\noindent
where $\mathcal{B}(\cdot,\cdot)$ is the beta function and $\Gamma(\cdot)$ is the gamma function.

\item
The Student's t distribution has a power-law tail. Let $X\sim t(\nu)$, we have:
\begin{equation*}
    f_X(x) = \frac{\Gamma(\frac{\nu+1}{2})}{\sqrt{\nu\pi}\Gamma(\frac{\nu}{2})} \left(1+\frac{x^2}{\nu}\right)^{-\frac{\nu+1}{2}}
\end{equation*}
The tail distribution is given by   
\begin{equation}\label{eq:MPMR:Student}  
\begin{aligned}
    f_X(x)& \approx \frac{\nu A^\nu}{x^{(\nu+1)}}  
    \\
    s^*& \approx
    A {\left(\frac{1 + n \nu}{1+\nu}\right)}^{\frac{1}{\nu}}
\end{aligned}
\end{equation}
where $A=\sqrt{\nu}\left(
\frac
{\Gamma(\frac{\nu+1}{2})}
{\sqrt{\pi}\Gamma(\frac{\nu+2}{2})}
\right)^{\frac{1}{\nu}}. $
    When $\nu>1$, we have:
\begin{equation}\label{eq:EMR:Student}
\begin{aligned}
    \mathbb{E}[S]
    &\approx
    An\mathcal{B}(n,1-\frac{1}{\nu})
    \\
    &\approx 
    A~ \Gamma(1-\frac{1}{\nu}) ~ n^\frac{1}{\nu} +
    \frac{A~ \Gamma(2-\frac{1}{\nu})}{2\nu} ~ n^{\frac{1}{\nu}-1}
\end{aligned}
\end{equation}

\end{enumerate}
%
%
\noindent
For these distributions we consider here, it is easy to check that the leading order dependence of \gls{mpmr} and \gls{emr} (when it exists) on the sample size $n$, as $n\rightarrow \infty$, is the same. This has been explicitly shown in the cases of the exponential and power-law distributions. Since the tail part of the size distribution of risk events can often be described by the Pareto-type model \cite{gomes_extreme_2015, nolde_extreme_2021}, we will focus on the case of power-law distribution in our tail risk measure study. It can be seen from \cref{eq:MPMR:Pareto,eq:EMR:Pareto,eq:MPMR:Student,eq:EMR:Student} that \gls{mpmr} and \gls{emr} scale with sample size $n$ with the \gls{se} $\eta=\frac{1}{\xi}$, or the reciprocal of the \gls{ti}, $\xi$.

\section{\gls{ti} estimator derived from sample-size scaling of \gls{mpmr}}
\label{sec_3}
\noindent
In this section, we explore the use of sample-size scaling of \gls{mpmr} to estimate \gls{ti} for power-law tail distributions and demonstrate that it provides a robust and accurate estimate of \gls{ti}. Given a sample size of $n$, we can perform a sub-sampling without replacement from $N$ raw observations to form a block of size $n$ and record its maximum as a \gls{bm} sample. We obtain the \gls{mpmr} at this block size $n$ by estimating the mode of $M$ sub-sampled \gls{bm}s using the mean-shift algorithm \cite{lee_finding_2021}, whose convergence in $\mathbb{R}^1$ is proven \cite{aliyari_ghassabeh_convergence_2013}. The hyper-parameter bandwidth used in the algorithm is calculated following the rule of thumb in \cite{dhaker2021beta}. The \gls{emr} at this block size $n$ is simply estimated by taking the sample mean on the $M$ sub-sampled \gls{bm}s. 
\\\\
With a series of \gls{mpmr} and \gls{emr} at different $n$, we fit a simple linear regression model to $\log\left(\text{MPMR}\right) \sim \log(n)$ and to $\log\left(\text{EMR}\right) \sim \log(n)$. We then obtain the estimate of \gls{se} $\hat{\eta}$ as the slope of the linear fit. Note that instead of fitting $\log\left(\text{EMR}\right) \sim \log(n)$, we can use all the raw BMs to fit a linear regression model to $\log\left(\text{BM}\right) \sim \log(n)$ and obtain the same result. The estimated \gls{ti} $\hat{\xi}$ is given as the reciprocal of $\hat{\eta}$. 
\\\\
During the sub-sampling step, we set the lower limit for $n$ as $n=1$, and the upper limit as $n=N(1-1/e)$. This is to take up a reasonably wide range of $n$ for regression purposes while avoiding \gls{bm} estimates at the same $n$ being too concentrated. Considering that we are handling mostly time-series data and the tail distribution is highly non-Gaussian, we selected sub-sampling instead of bootstrapping or upstrapping \cite{politis_asymptotic_2001, crainiceanu_upstrap_2020}. Following Ref. \cite{wilcox2010bootstrap}, we set the number of sub-samples $M$ to 600 at each $n$.
\\\\
We conduct simulations to compare our \gls{ti} estimators based on \gls{mpmr} to several well-known \gls{ti} estimators \cite{munasinghe2020fast, fedotenkov_review_2020}, including Hill's estimator \cite{hill1975simple}, least squares estimator \cite{zaher2014estimation}, weighted least squares estimator \cite{nair2013fundamentals}, method of percentiles estimator \cite{bhatti2018efficient}, method of modified percentiles estimator \cite{bhatti2018efficient}, maximum likelihood estimator \cite{newman2005power}, method of geometric percentiles estimator \cite{bhatti2018efficient}, and method of moments estimator \cite{brazauskas2000robust,rytgaard1990estimation}. We use Student's t as the underlying distribution to generate raw samples of various \gls{ti}s and sample sizes. Only the top $10\%$ of raw samples are picked as effective tail samples of size $N$ for the estimators to work on. From \cref{eq:MPMR:Student,eq:EMR:Student}, the true \gls{ti} $\xi$ is equal to the degree of freedom $\nu$. For every combination of effective sample size $N$ and $\nu$, we run $1009$ simulations to get the mean, standard deviation and mean absolute percentage error using outcomes from the estimators considered here.
\begin{figure}
    \centering
    \begin{subfigure}[b]{.9\textwidth}
        \centering
        \includegraphics[width=1.1\textwidth]{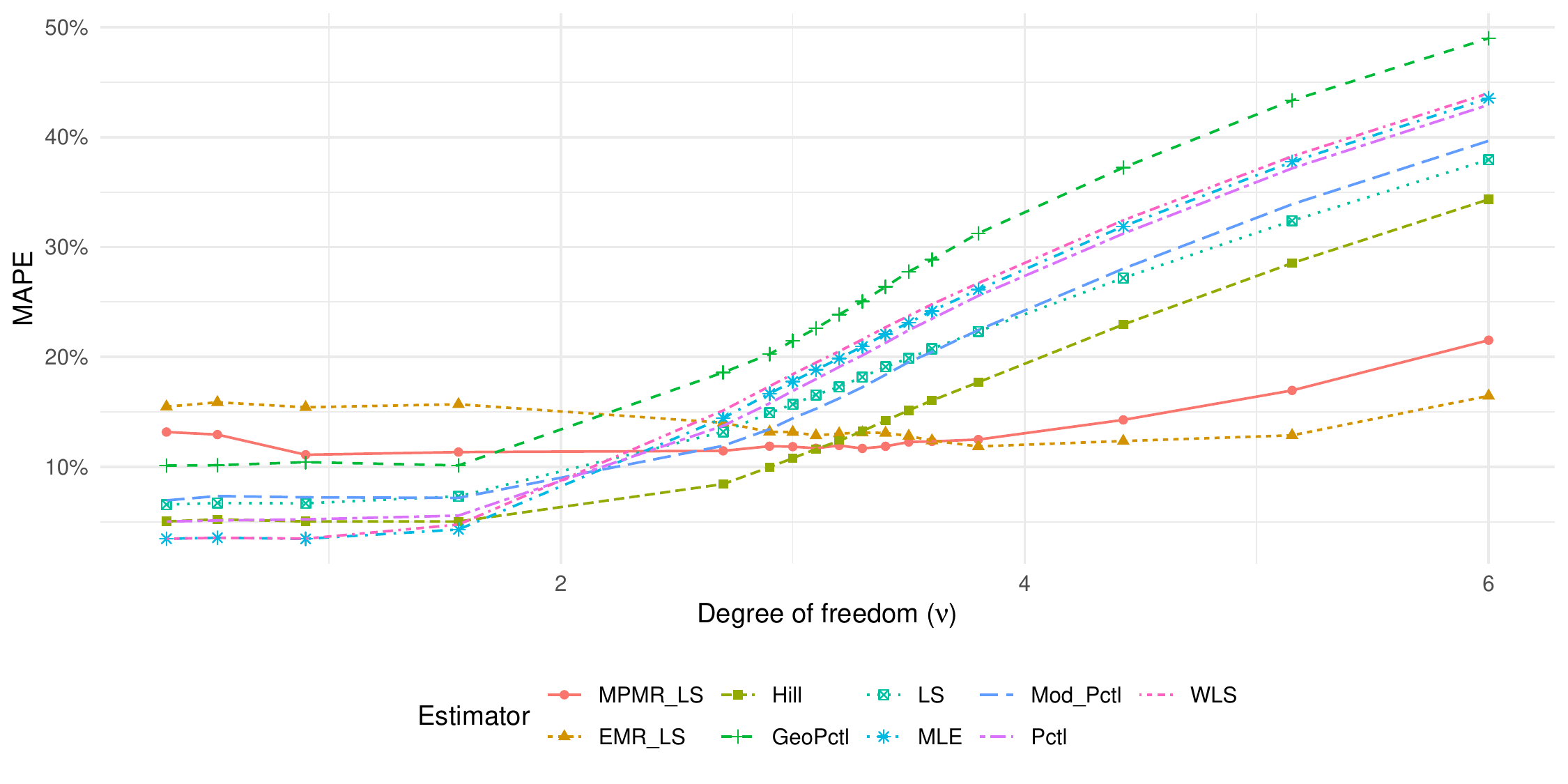}
        \caption{Mean absolute percentage error of \gls{ti} estimators at various true \gls{ti}s, with effective sample size = $1000$.}
        \label{fig:TI_MAPE}
    \end{subfigure}
    \begin{subfigure}[b]{.9\textwidth}
        \centering
        \includegraphics[width=1.1\textwidth]{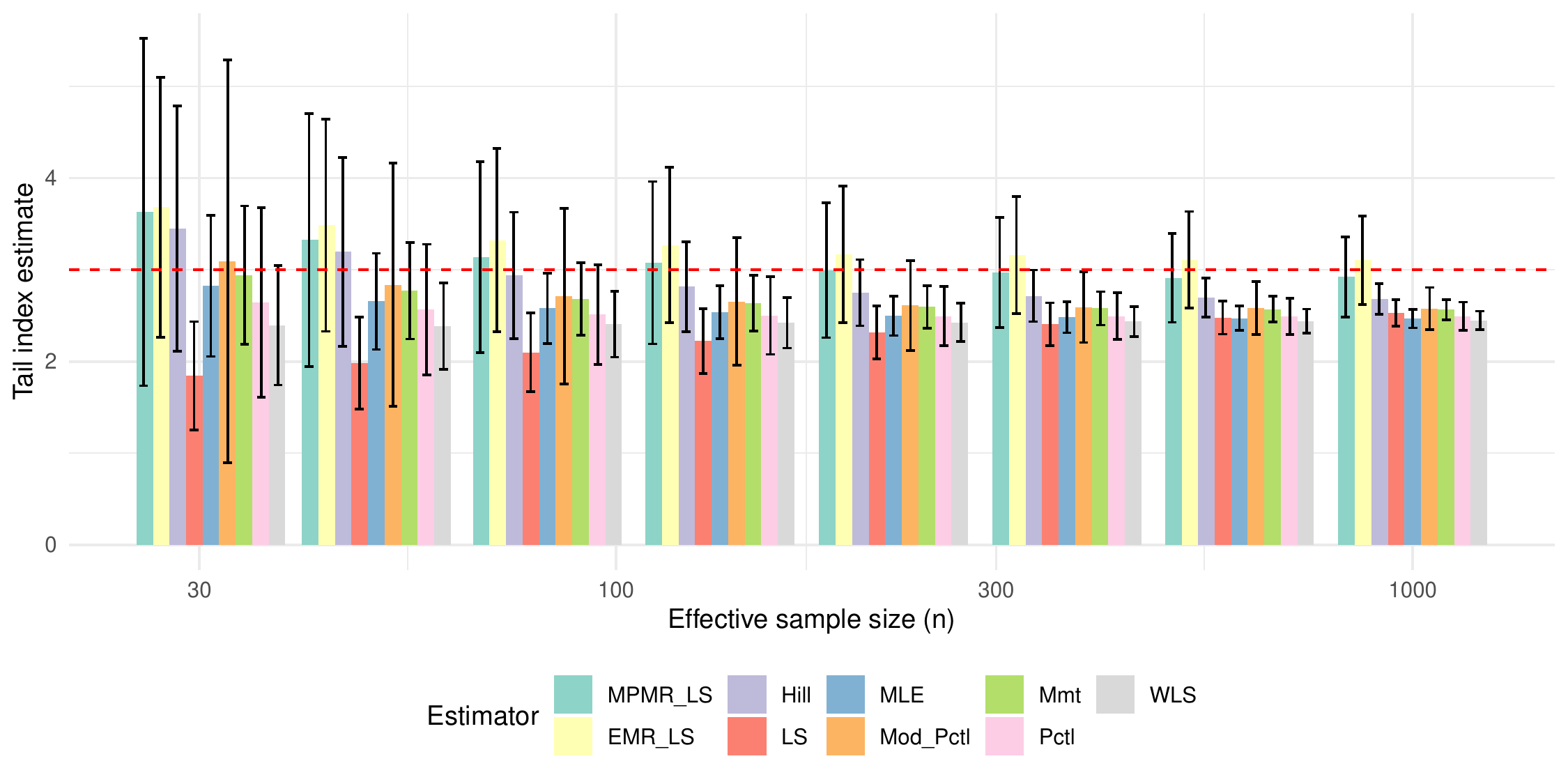}
        \caption{Mean and standard deviation of \gls{ti} estimators at various effective sample sizes, with $\nu=3$.}
        \label{fig:TI_n}
    \end{subfigure}
    \caption{Performances of \gls{ti} estimators, applied on top $10\%$ observations from Student's t distribution.}
    \label{fig:TI}
\end{figure}
\\\\
Comparisons are charted in \cref{fig:TI}. Estimators based on \gls{mpmr} and \gls{emr} have good bias-variance trade-offs for various $\nu$ as shown in \cref{fig:TI_MAPE}. We further fix $\nu=3$ (to match the cubic law found in stock returns \cite{gabaix_theory_2003}) and show the mean and standard deviation of TI estimates at various effective sample sizes in \cref{fig:TI_n}. The \gls{mpmr} and \gls{emr} estimates show consistently low biases. Note that for financial markets' daily returns, the tail sample (taking 10\% of the top profit and loss values from the raw data sample of around 252 return values in a year) has around $13$ effective observations of losses. In such small sample scenarios, \gls{mpmr} or \gls{emr} is a relatively more robust choice. The RBM estimator from Ref. \cite{wager_subsampling_2014} also gives good estimates but requires expensive computation.

\section{Stylized facts of \gls{mpmr}/\gls{emr} for describing the natural hazards}
\label{sec_4}
\noindent
In this section, we apply the proposed \gls{mpmr} and \gls{emr} to analyze the tail risks associated with natural hazards. Extreme value analysis is widely accepted to analyze, understand, and predict natural hazard events \cite{davison2015statistics, haigh2019advances}. The selected examples of natural hazards have significant impacts on human activities, and the associated risks have gradually been quantified and priced in financial markets. There are also ample observation data on these natural hazards over a long history; this makes the statistical analysis of natural hazards a good test for tail risk measures,

\subsection{Earthquake}
\begin{figure}
    \begin{subfigure}[b]{.9\textwidth}
        \centering
        \includegraphics[width=1.1\textwidth]{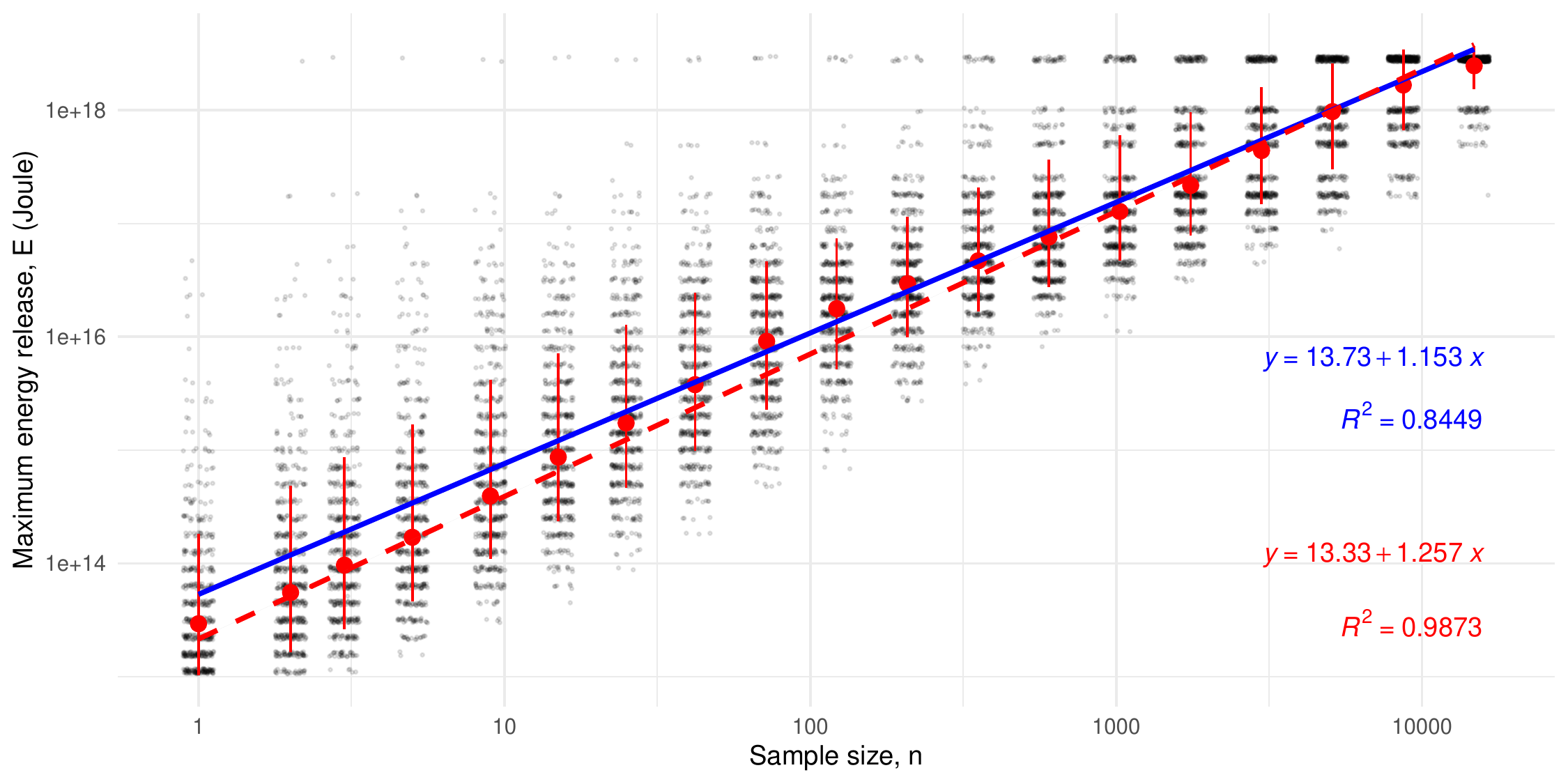}
        \caption{Maximum energy release among $n$ earthquake occurrences. Red bars cover mean $\pm$ sd. Red dots mark mode estimates. The blue solid line fits all data and the red dashed line fits mode estimates.}
        \label{fig:BM_earthquake}
    \end{subfigure}
    \begin{subfigure}[b]{.9\textwidth}
        \centering
        \includegraphics[width=1.1\textwidth]{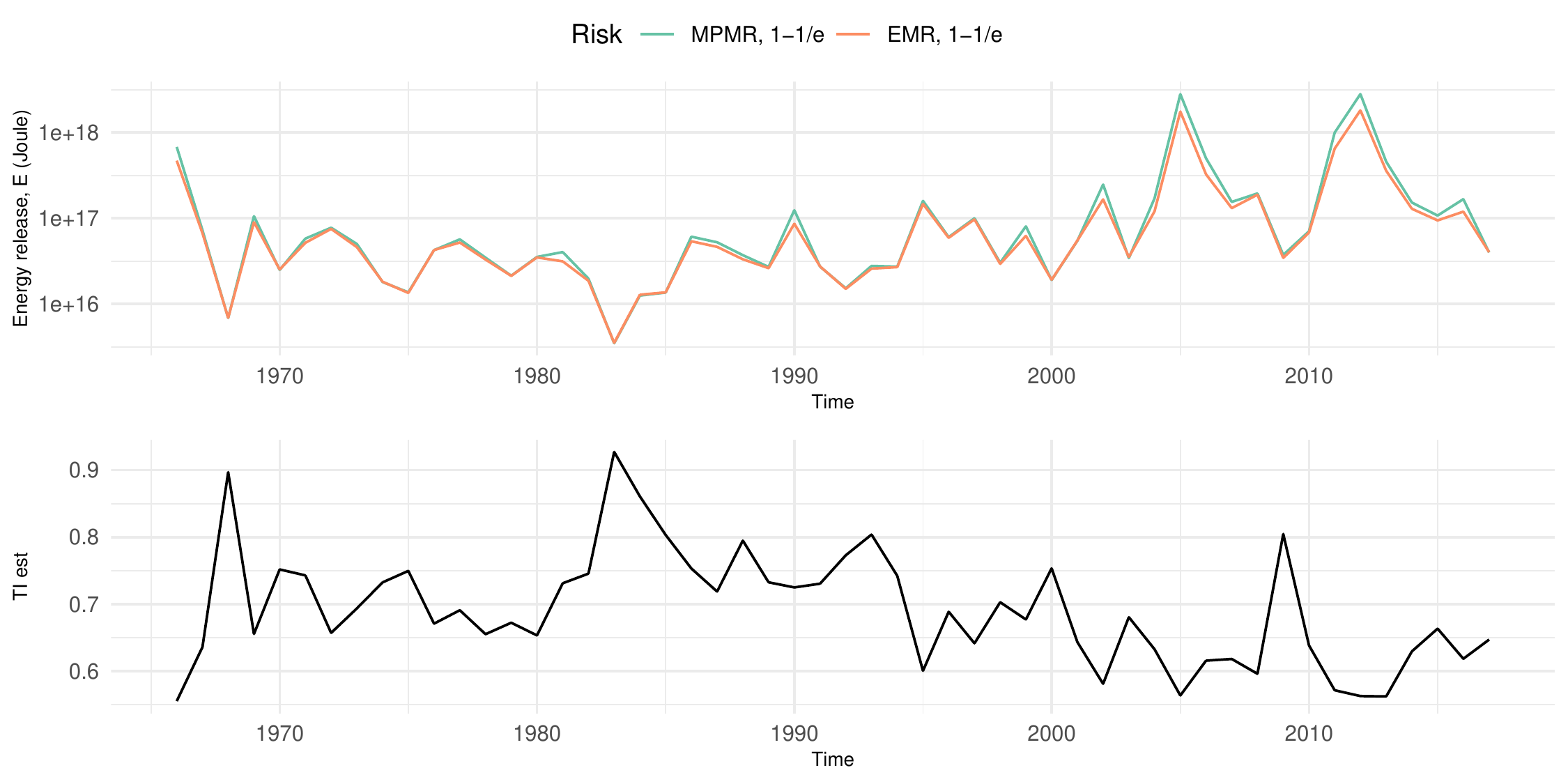}
        \caption{Time series of earthquake \gls{mpmr}, \gls{emr}, and \gls{ti}, using observations within each year.}
        \label{fig:ts_earthquake}
    \end{subfigure}
    \centering
    \caption{}
    \label{fig:earthquake}
\end{figure}

\noindent
We first consider earthquake statistics. The size-frequency relation of earthquakes is well described by a power law, known as the Gutenberg-Richter law, which states that the probability of an earthquake with a size greater than $S$ is proportional to $S^{-b}$, or $\mathbb{P}(\text{Size}>S)\propto S^{-b}$. The original GR law is stated as a frequency-magnitude relation, where the magnitude is the earthquake size on a log scale. We can also write the GR law as an energy-frequency relation. The conversion between the energy release and the moment magnitude is $M_w=(\frac{2}{3})\log_{10}(E)-3.2$ (so $S \propto E^{2/3}$; this gives rise to the energy-frequency relation, $\mathbb{P}(\text{Energy}>E)\propto E^{-B}$, where the exponent $B=\frac{2}{3}b$). For our analysis, we use a historical dataset from the US Geological Survey, which contains all $23,412$ significant earthquakes with a magnitude of $5.5$ or higher from $1965$ to $2016$. 
\\\\
In \cref{fig:BM_earthquake} we show the sub-sampled \gls{bm}s of the earthquake energy release together with the \gls{mpmr} (the most probable maximum energy, obtained from mode estimate) and \gls{emr} (the expected maximum energy, estimated as the sample mean) at various sample sizes $n$. The log-log plots of \gls{mpmr} versus the sample size show a scaling relation: $E_{\text{MPMR}}(n) \approx 2.15 \times 10^{13} ~ n^{\eta}$ Joules, where the \gls{se}, $\eta \approx 1.257$. From this scaling relation, we can expect the energy of the maximum earthquake in one year (we use $n=450$, which is the average number of earthquakes in one year in the data set) $\approx 4.67 \times 10^{16}$ Joules (the magnitude $M_w \approx 7.91$), the energy of the maximum earthquake in $10$ years $\approx 8.44 \times 10^{17}$ Joules ($M_w \approx 8.75$), and the energy of the maximum earthquake in $100$ years $\approx 1.53\times 10^{19}$ ($M_w \approx 9.58$), which is close to the magnitude of the great $1960$ Chilean earthquake. The \gls{se} of $\eta = 1.257$ gives rise to the tail index $\text{TI},\text{B}=0.79$ and $b=1.19$. This $B$-value is quite close to the $B$-value ($\approx 0.7$) obtained from a quasi-static model of earthquakes in an earthquake zone with a pre-existing fault \cite{chen1997}. The \gls{se} obtained from the \gls{emr} is slightly smaller with $\eta \approx 1.15$ due to higher estimates obtained for the \gls{emr} for small $n$. The \gls{mpmr} and \gls{emr} converge as the sample size increases. 
\\\\
In \cref{fig:ts_earthquake} we show the time series of \gls{mpmr}, \gls{emr}, and \gls{ti} estimates, using aggregated observations within each year. Since the counts of observations are changing, we set the sample size as $1-\frac{1}{e}$ of the counts in each aggregation. The maximum energy estimates based on \gls{mpmr} and \gls{emr} are very close for these sample sizes. Note that the estimated \gls{ti} ($B$-value) based on earthquake statistics within a year varies from year to year in the range of $0.56$ to $0.93$ and is shown negatively correlated with the \gls{mpmr}. 

\subsection{Tsunami}
\begin{figure}
    \begin{subfigure}[b]{.9\textwidth}
        \centering
        \includegraphics[width=1.1\textwidth]{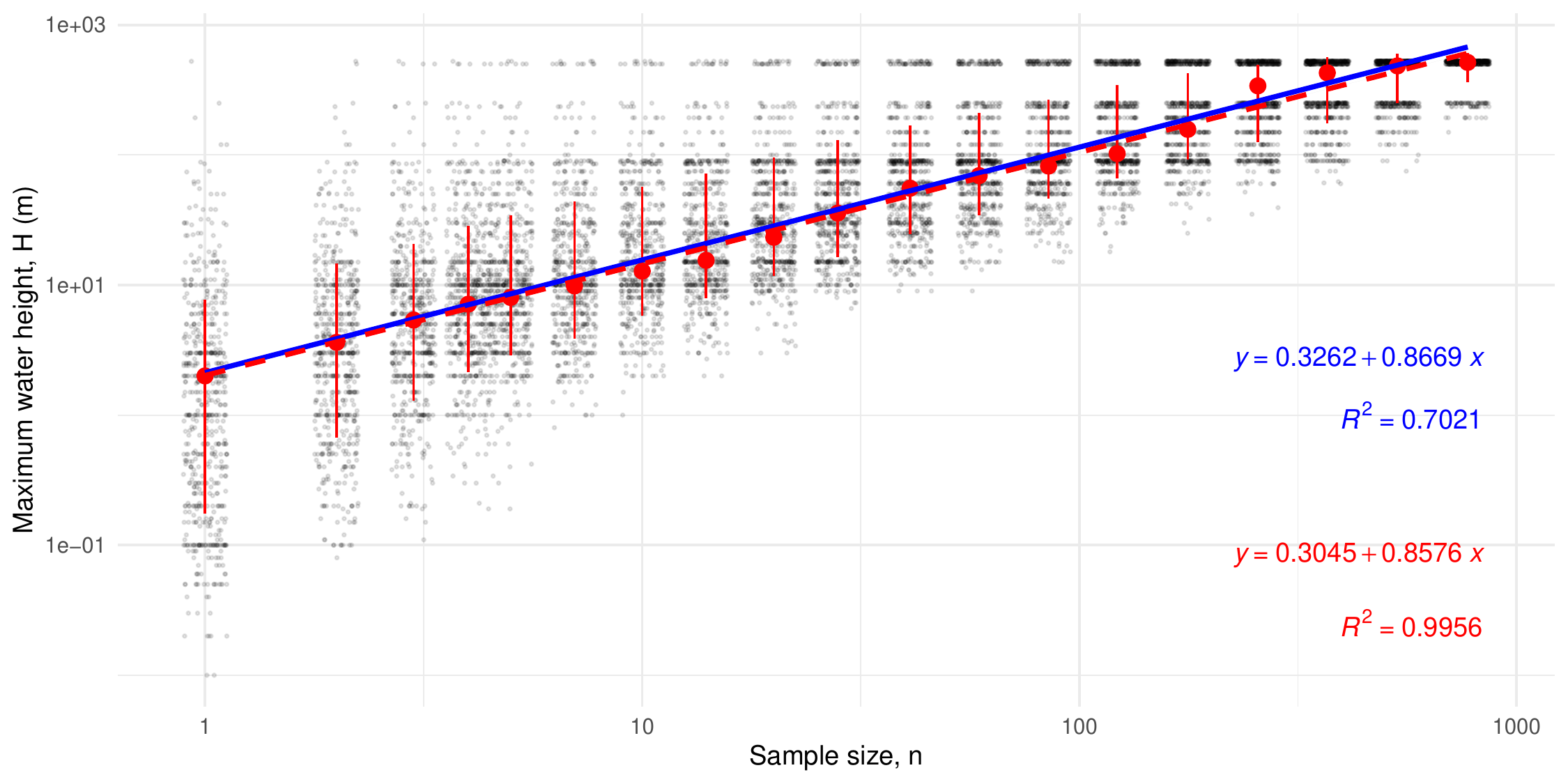}
        \caption{Maximum water height among $n$ tsunami occurrences. Red bars cover mean $\pm$ sd. Red dots mark mode estimates. The blue solid line fits all data and the red dashed line fits mode estimates.}
        \label{fig:BM_tsunami}
    \end{subfigure}
    \begin{subfigure}[b]{.9\textwidth}
        \centering
        \includegraphics[width=1.1\textwidth]{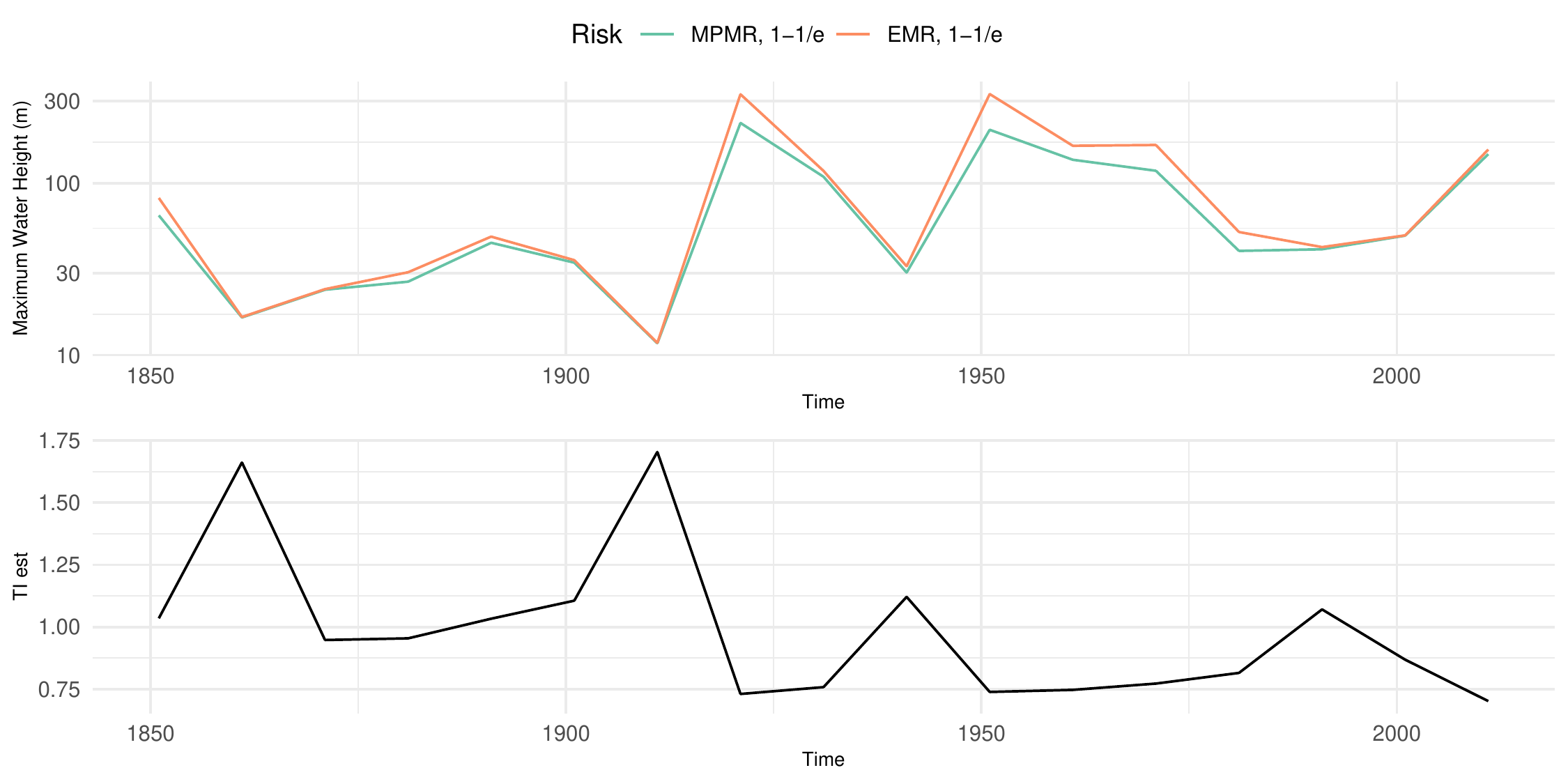}
        \caption{Time series of tsunami \gls{mpmr}, \gls{emr}, and \gls{ti}, using observations every ten years.}
        \label{fig:ts_tsunami}
    \end{subfigure}
    \centering
    \caption{}
    \label{fig:tsunami}
\end{figure}

\noindent
In the second example, we turn to Tsunami statistics. We use the data from NGDC/WDS Global Historical Tsunami Database \cite{arcos2019past}. 
\\\\
In \cref{fig:BM_tsunami} we show the sub-sampled \gls{bm}s together with the \gls{mpmr} and \gls{emr} estimates of the maximum water height in tsunami events. The power-law scaling works well for the dependence of \gls{mpmr} and \gls{emr} on the number of tsunamis in the sample; the dependence is estimated as $n^{0.86}$. This corresponds to the \gls{ti}, $\xi \approx 1.16$ for the maximum water height distribution. The statistics of tsunami sizes should be related to the corresponding earthquake statistics. Our empirical analysis suggests $b \approx 1.19$, which is close to the tail index for tsunami water height $\xi \approx 1.16$. This suggests the relation between the maximum water height $H$ and the earthquake magnitude: $M_w \approx \log(H) + \text{const}$, consistent with the empirical analysis of Abe \cite{abe1979}. 
\\\\
In \cref{fig:ts_tsunami} we show the time series of \gls{mpmr}, \gls{emr}, and \gls{ti} estimates, using aggregated observations every ten years. We set the sample size as $1-\frac{1}{e}$ of the number in each aggregation. The \gls{mpmr} and \gls{emr} are very close in most cases but there are discrepancies at some time intervals due to very small raw sample sizes.

\subsection{Rainfall}
\begin{figure}
    \begin{subfigure}[b]{.9\textwidth}
        \centering
        \includegraphics[width=1.1\textwidth]{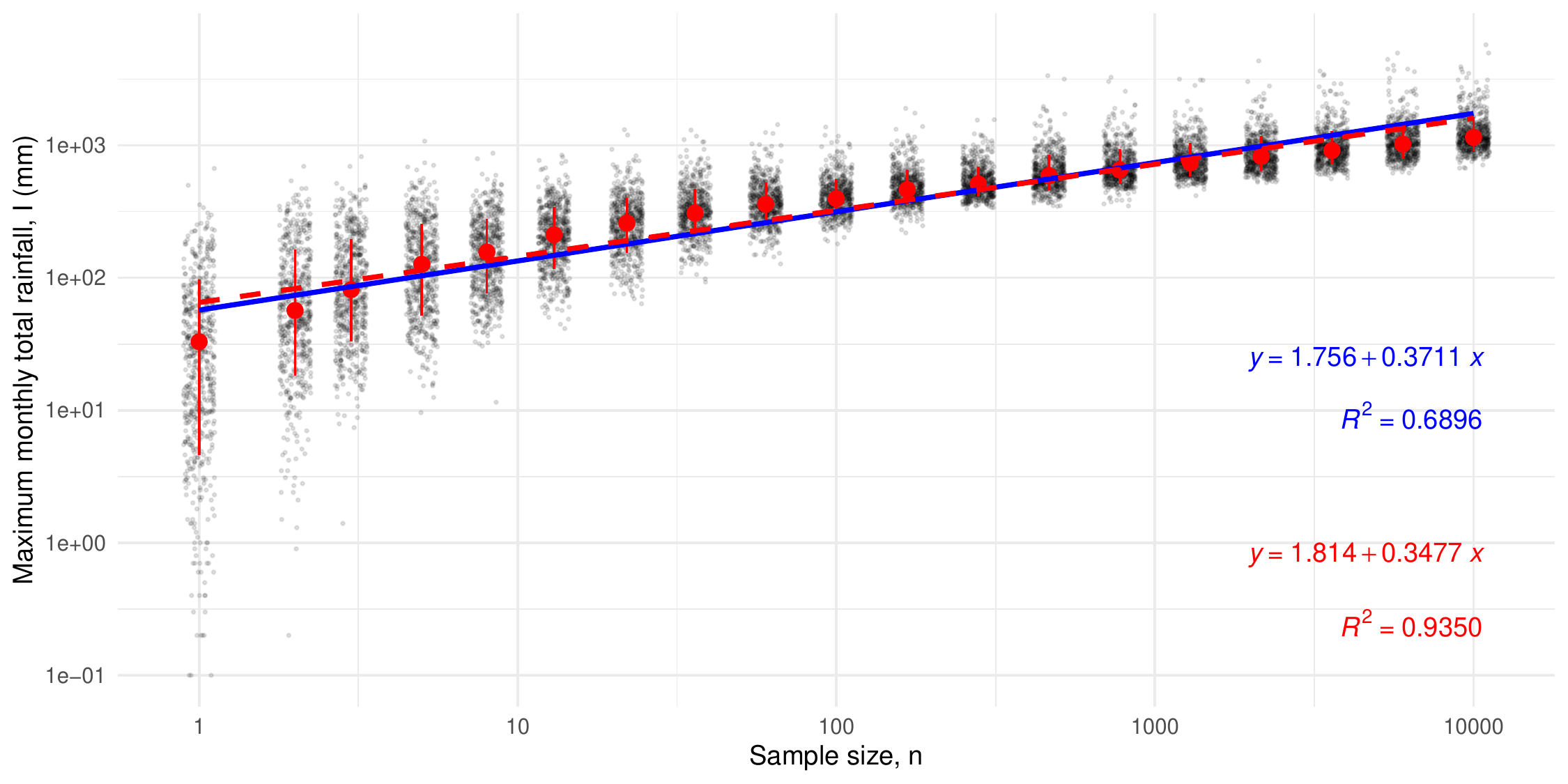}
        \caption{Maximum monthly accumulated rainfall (monthly total precipitation) among $n$ months. Red bars cover mean $\pm$ sd. Red dots mark mode estimates. The blue solid line fits all data and the red dashed line fits mode estimates.}
        \label{fig:BM_rainfall}
    \end{subfigure}
    \begin{subfigure}[b]{.9\textwidth}
        \centering
        \includegraphics[width=1.1\textwidth]{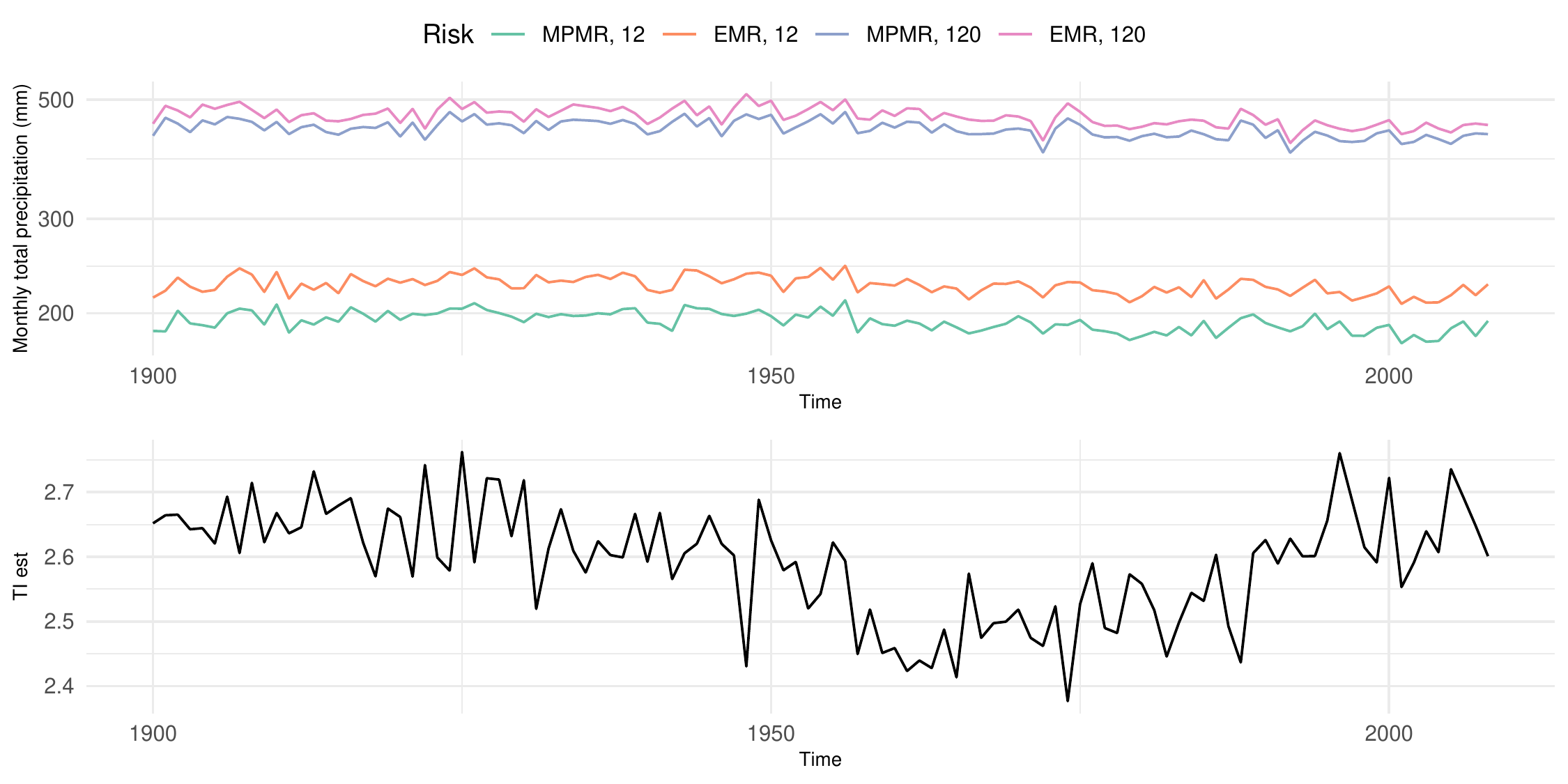}
        \caption{Time series of rainfall \gls{mpmr}, \gls{emr}, and \gls{ti}, using observations within each year.}
        \label{fig:ts_rainfall}
    \end{subfigure}
    \centering
    \caption{}
    \label{fig:rainfall}
\end{figure}

\noindent
In the third example, we investigate the maximum monthly accumulated rainfall (monthly total precipitation) in monthly rainfall samples of various sizes. We use the data from the University of Delaware \cite{matsuura2018terrestrial} which contains global monthly rainfall data from $1900$ to $2008$ ($1308$ months) at $85,794$ locations. We remove the observations with $0$ monthly total precipitation to focus on flood instead of drought hazards.
\\\\
In \cref{fig:BM_rainfall} we show the sub-sampled \gls{bm}s together with the \gls{mpmr} and \gls{emr} estimates of maximum monthly total precipitation for a different number of months in the sample (sample size). To avoid the spatial-temporal proximal dependence that might deteriorate the i.i.d. assumption, we set the upper bound for the sample size as $n=10,000$, which makes our samples sparse in space and time. From the figure, the maximum precipitation scales with $n$, the sample size, approximately $n^{0.35}$. This gives rise to the TI estimate of approximately $2.88$. 
\\\\
In \cref{fig:ts_rainfall} we show the time series of \gls{mpmr}, \gls{emr}, and \gls{ti} estimates, using aggregated observations within each year. In each aggregation, we have $12\times85,794=1,029,528$ observations, on which we sub-sample sparsely for either $n=12$ or $n=120$. The \gls{mpmr} and \gls{emr} are close and the values are quite stable for more than $100$ years. The \gls{ti} shows some cyclical fluctuation, perhaps reflecting a certain time-series correlation. This remains to be explored in future research.  

\section{Stylized facts of \gls{mpmr}/\gls{emr} on financial market risk}
\label{sec_5}
\begin{figure}
    \hspace*{-.1\textwidth}%
    \subcaptionbox{}{
        \includegraphics[width=.55\textwidth]{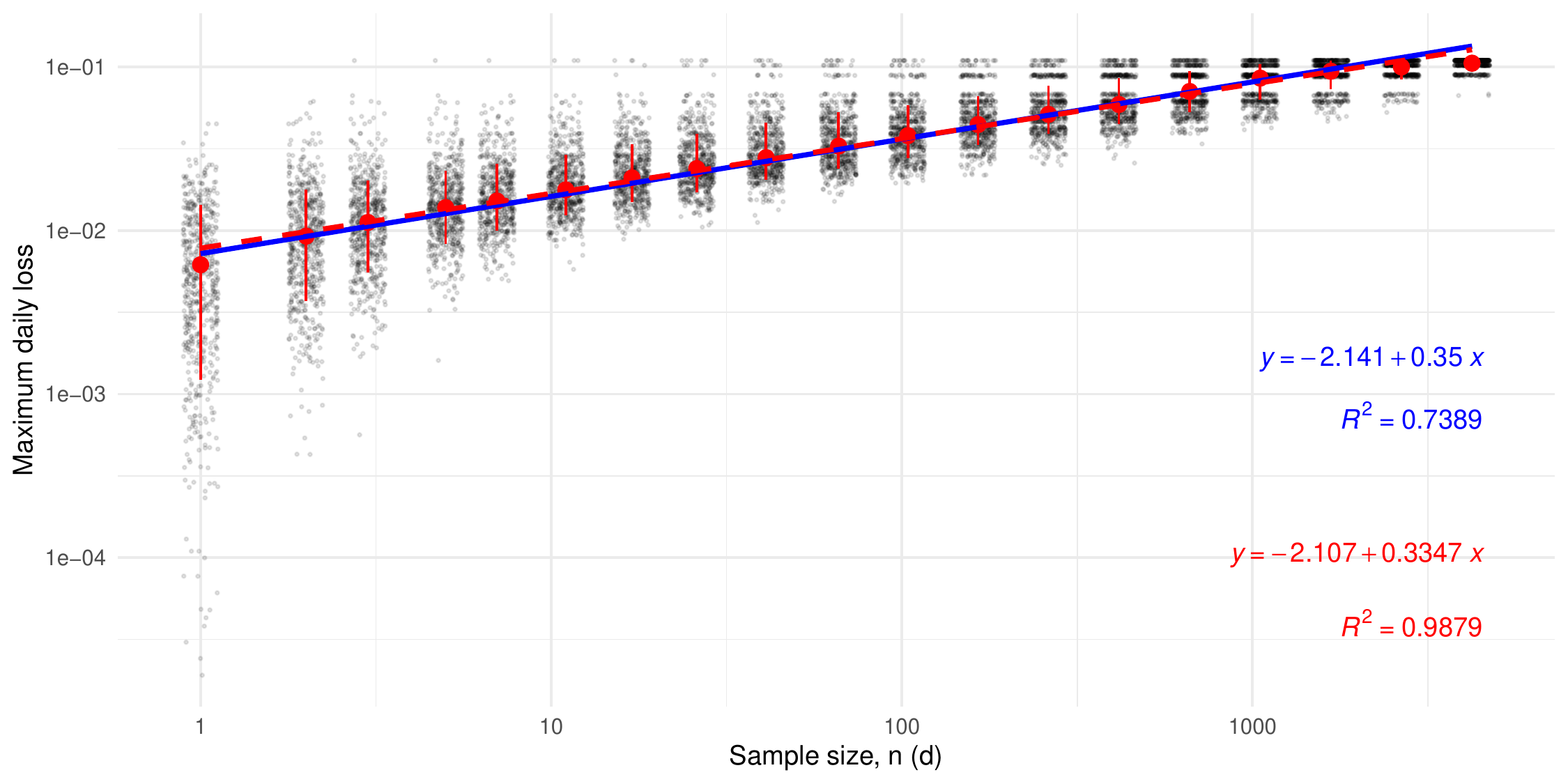}
        \label{fig:BM_SP500}
    }
    \hfill
    \subcaptionbox{}{
        \includegraphics[width=.55\textwidth]{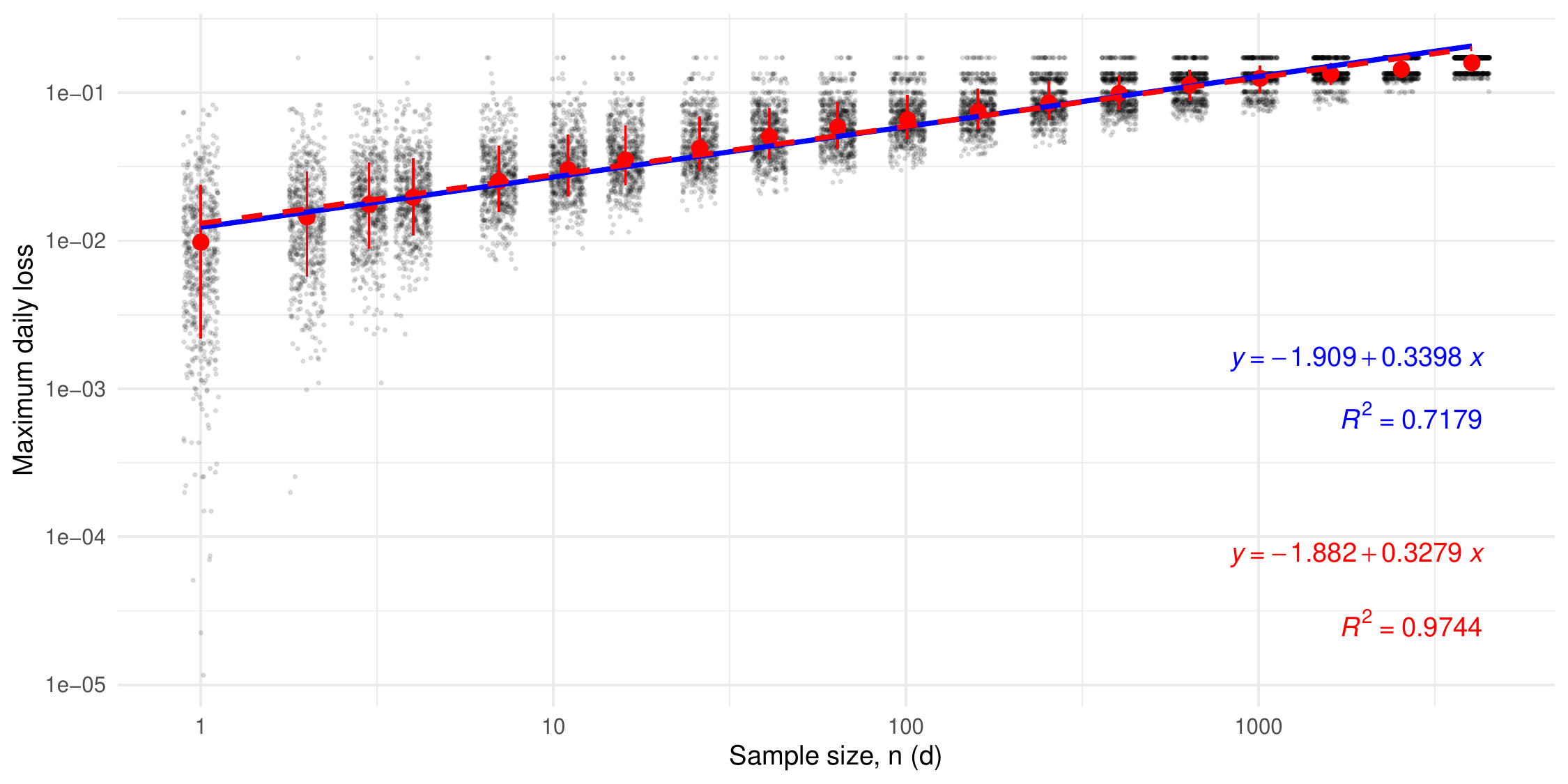}
        \label{fig:BM_HSI}
    }
    \hspace*{-.1\textwidth}%
    \\
    \hspace*{-.1\textwidth}%
    \subcaptionbox{}{
        \includegraphics[width=.55\textwidth]{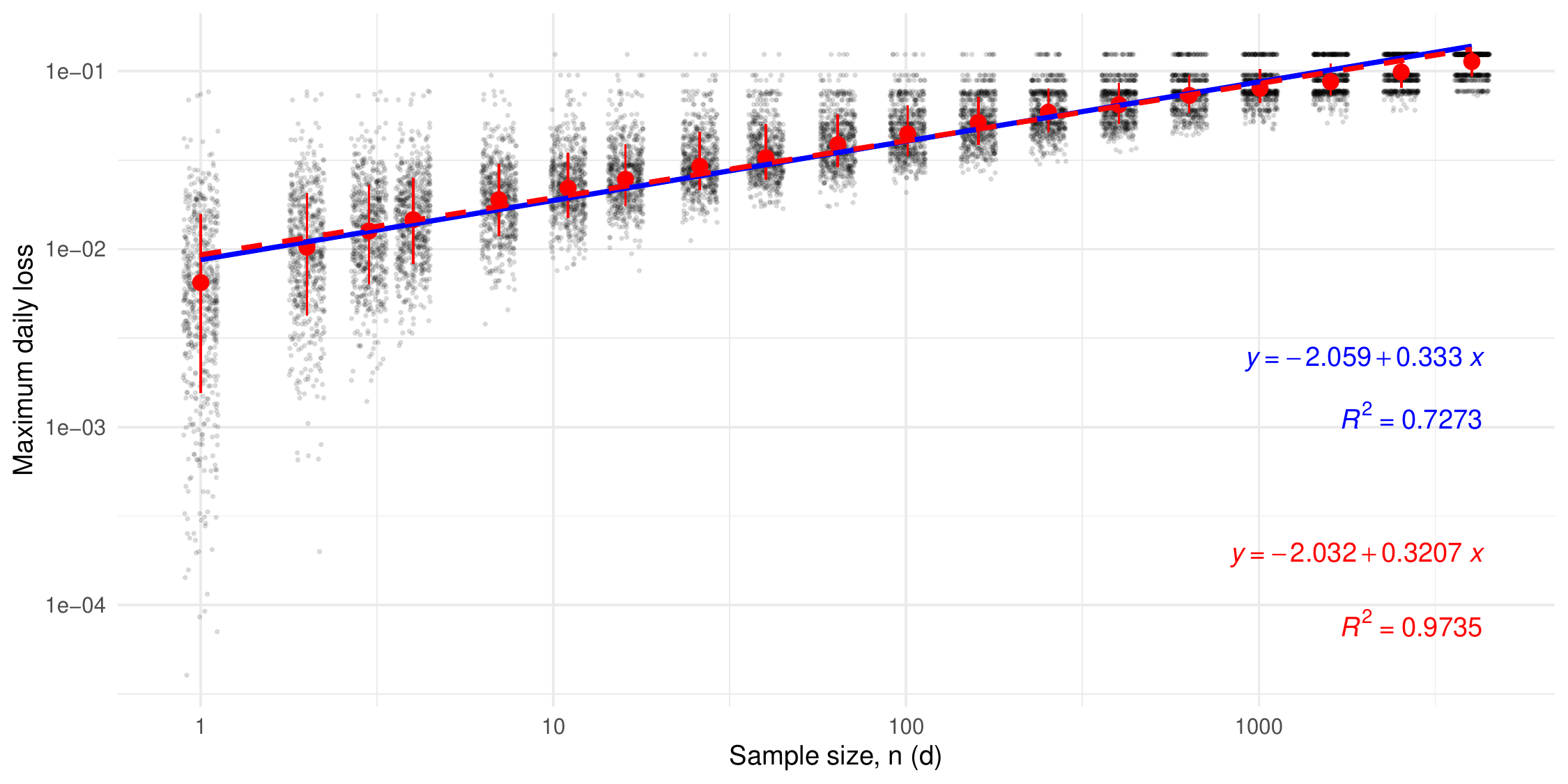}
        \label{fig:BM_NKY}
    }
    \hfill
    \subcaptionbox{}{
        \includegraphics[width=.55\textwidth]{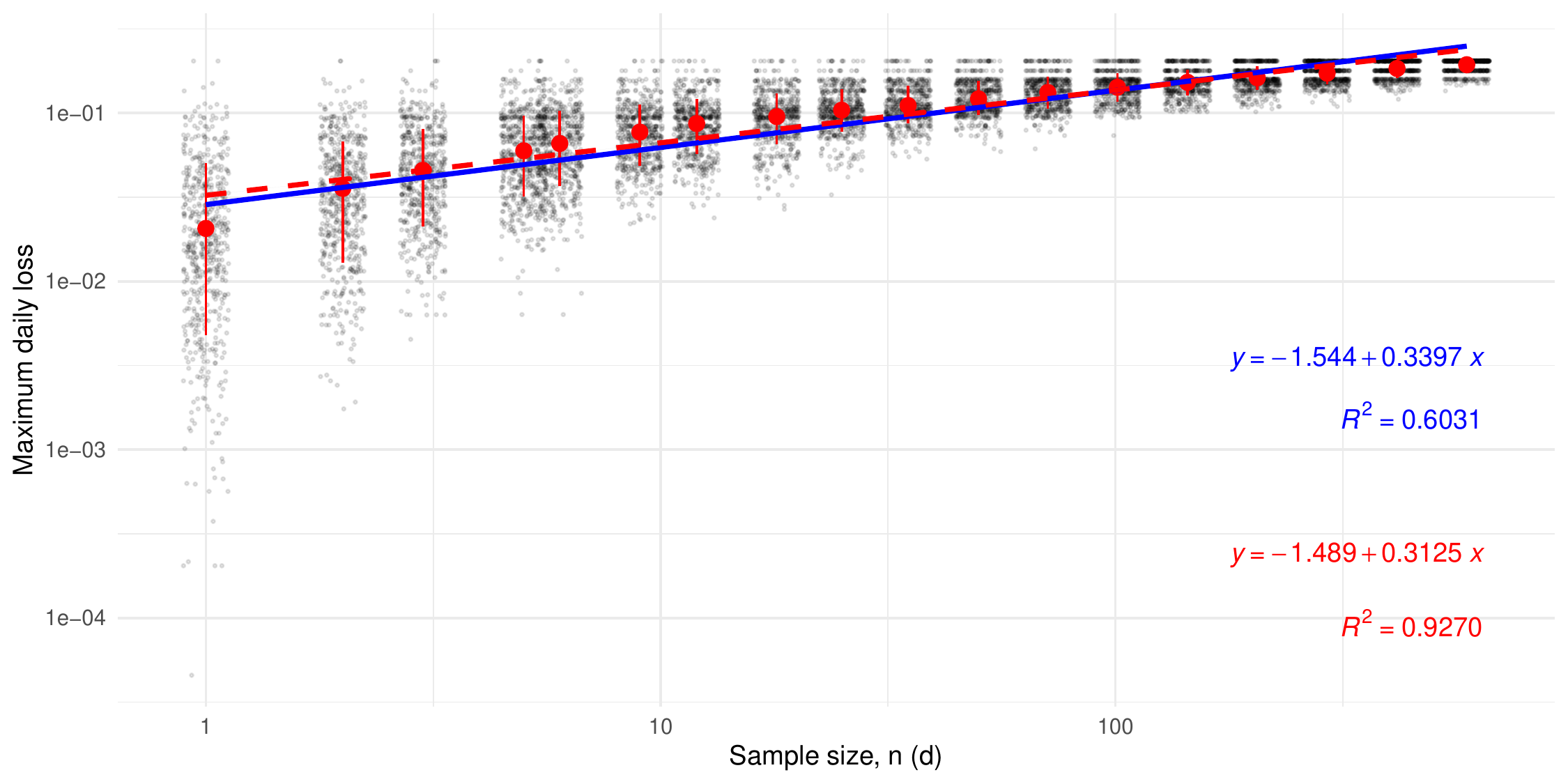}
        \label{fig:BM_BTC}
    }
    \hspace*{-.1\textwidth}%
    \caption{Maximum daily loss among $n$ days, for (a) SP500, (b) HSI, (c) NKY, and (d) BTC. Red bars cover mean $\pm$ sd. Red dots mark mode estimates. The blue solid line fits all data and the red dashed line fits mode estimates.}
    \label{fig:BM_finloss}
\end{figure}

\noindent
In this section, we focus again on financial risks and apply the proposed \gls{mpmr} and \gls{emr} measures for estimating market risks. We use various market indices and bitcoin price series for our analysis; these include SP500 (from 1970-01-02 to 2022-07-29), HSI (from 1969-11-24 to 2022-06-30), NKY (from 1970-01-05 to 2022-06-30), and BTC (from 2017-09-02 to 2022-08-03). We consider the percentage log loss, $\text{loss}=\max(-\text{log\_return}, 0)$ (the loss is set to $0$ for positive returns). To focus on the tail-risk behavior, we remove zero loss from the raw samples.
\\\\
In \cref{fig:BM_finloss} we show the sub-sampled \gls{bm}s together with the \gls{mpmr} and \gls{emr} estimates of the maximum daily loss for a different number of days in the sample (sample size). From the figure, we can see that the maximum loss scale with $n$, the sample size approximately as $n^{\eta}$. The scaling exponent $\eta$, estimated using either \gls{mpmr} or \gls{emr}, is close to $0.33$ for all the indices and BTC we consider. This gives rise to the \gls{ti} estimate of approximately $3$, strongly supporting the validity of the cubic law.  The scaling relation can also be used to estimate the risk in a longer time frame than the time frame of the available data. For the SP500 index, the maximum daily loss is about $1.69\%$ on average when $T=10$ days, about $4.97\%$ when the time frame is $1$ year (corresponding to $T=252$ trading days in a year), and about $10.74\%$ when the time frame is $10$ years. The scale invariance gives a reasonable estimate of about $23.21\%$ when extrapolating to the time frame of $100$ years. 

\begin{figure}
    \includegraphics[width=1.1\textwidth]{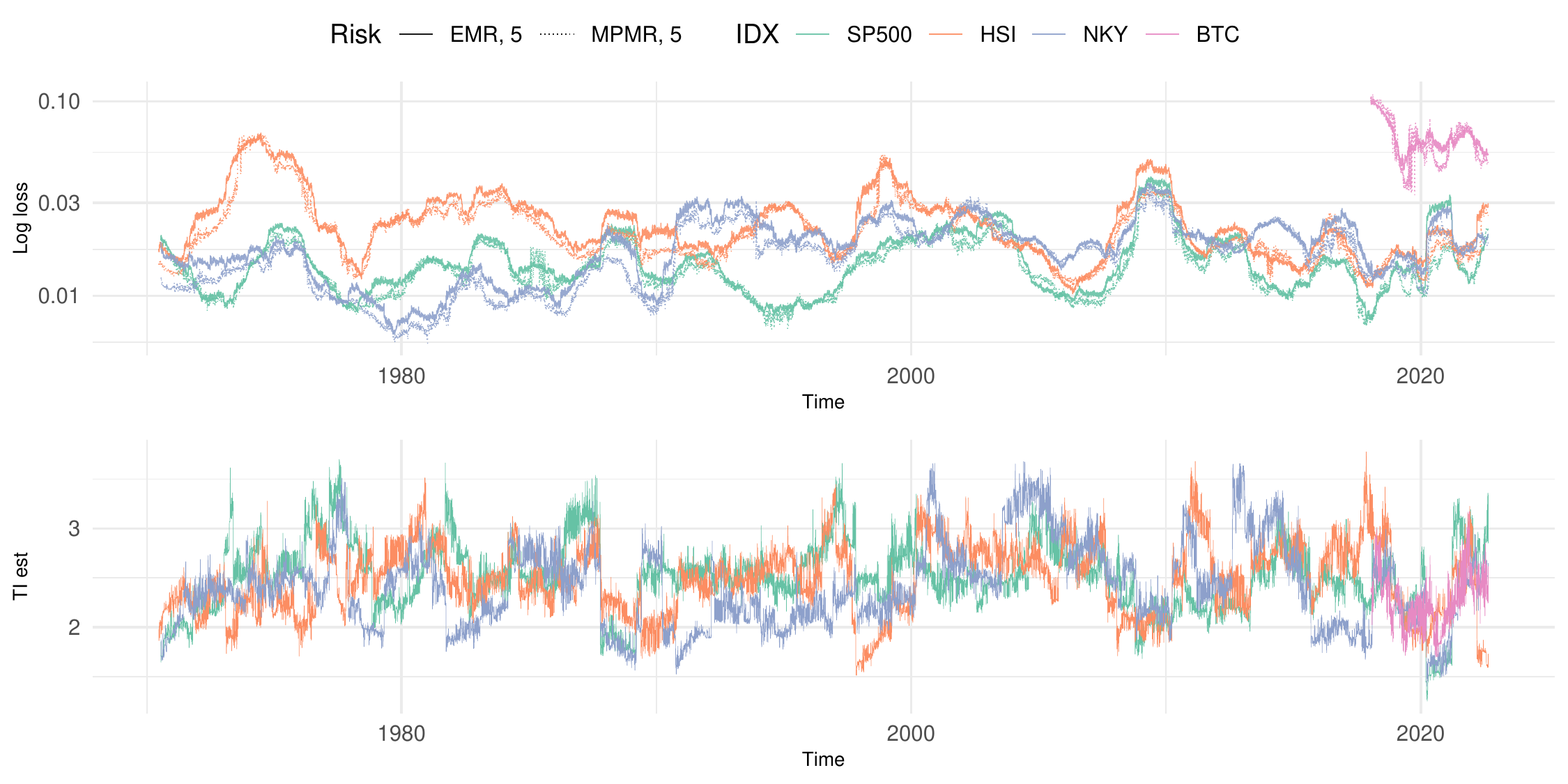}
    \centering
    \caption{Time series of financial loss \gls{mpmr}$_{5d}$, \gls{emr}$_{5d}$, and \gls{ti}, using rolling 365d window.}
    \label{fig:ts_finloss}
\end{figure}

\noindent
\\\\
In \cref{fig:ts_finloss} we show the time series of \gls{mpmr}, \gls{emr}, and \gls{ti} estimates, using lagged rolling $365$-day windows. Inside each window, we sub-sample with $n=5$. The values of \gls{mpmr} and \gls{emr} are very close, even for a small sample size of $5$. The time series also show that the risk metrics of different assets rise together (indicating increasing correlations and worse diversification) in a stressed market. The \gls{ti} shows strong oscillation in these $365$-day rolling window estimates. Its dependence on the market condition is worth further investigation.

\section{Conclusion}
\label{conclusion}
\noindent
In this paper, we propose tail risk measures based on the maximum value of risk for a risk event within a time frame. We show the analytical forms of \gls{mpmr} and \gls{emr} for several well-known underlying distributions. For the case of power-law tail distribution, the risk measure is shown to scale with the sample size with the scaling exponent $\eta$ equal to the reciprocal of the tail index $\xi$: $\eta=1/\xi$. By estimating $\eta$ we can construct an estimator for $\xi$. We have shown that such an estimator provides low-bias and consistent estimates (particularly for higher values of \glspl{ti}, which are difficult to estimate directly but are widely seen in practice). We have shown robust sample-size scaling of the risk measures obtained from a wide range of data sources: from earthquakes, tsunamis, and rainfall, to financial market risks. This scale invariance helps relate the risks in different time frames and allows for using extrapolation to estimate the risk in a long time frame. Our analysis of the financial market risk using \gls{mpmr} and \gls{emr} further validates the cubic law for the tail distribution of market risk; in particular, we have shown that the cubic law also applies to the cryptocurrency market. The analysis of the earthquake, tsunami, and rainfall data gives rise to a more intuitive measure of risks associated with such natural hazards in various time frames. These analyses also help to confirm some of the earlier theoretical and empirical results. The analysis of rainfall data gives stable estimates of \gls{mpmr} and \gls{emr} spanning more than $100$ years. Such assessments are important for managing long-term climate risks associated with extreme rainfall.

\section*{Acknowledgment}
We thank Ms. Anzhi Liao for her help in conducting some simulations. This research is partly supported by a Ministry of Education academic research grant on Outbreak Resilience, Forecasting, and Early Warning.
\\\\


\bibliographystyle{elsarticle-num-names} 
\bibliography{0_main}





\end{document}